\documentclass[prd,preprint,showpacs,groupedaddress]{revtex4-1}
\usepackage{amsmath}
\usepackage{amsfonts}
\usepackage{amssymb}
\usepackage{geometry}
\usepackage{natbib}
\usepackage[english]{babel}
\usepackage{indentfirst}
\usepackage{mathrsfs}
\usepackage{graphicx}

\begin{document}

  \setlength{\parindent}{2em}
  \title{The evolution of universe in the two-scalar theory }
  \author{Jia-Cheng Ding} \author{Ping Li} \author{Cong Li} \author{Qi-Qi Fan}
  \author{Jian-Bo Deng} \email[Jian-Bo Deng: ]{dengjb@lzu.edu.cn}
  \affiliation{Institute of Theoretical Physics, Lanzhou University, Lanzhou 730000, P. R. China}
  \date{\today}

  \begin{abstract}
  We generalize $f(R,T)$ gravity into the two-scalar theory that includes two independent scalar fields by the variational method, and we derive its field equations in Einstein frame using conformal transformation. Based on Friedmann equations and Raychaudhuri equation, with a consideration of the cosmic content as its perfect-fluid form, a further discussion leads us to an accelerated expanding condition of universe. In the two-scalar theory, universe has two states which are the accelerated expansion and decelerated contraction, and it has three stages during its evolution. The first and third stages are in the accelerated expanding state, and the second stage is in the decelerated contracting state. The third stage represents the present universe and it tends to become a dust universe.
  \end{abstract}

  \pacs{04.20.-q, 04.50.-h}

  \keywords{$f(R,T)$ gravity, two-scalar model, Raychaudhuri equation, the accelerated expanding universe}

  \maketitle

  \section{Introduction}

  Numerous and complementary cosmological observations show that our universe today is experiencing an accelerated expansion~\cite{riess1998ag,perlmutter1999s,de2000flat,schrabback2010evidence}.
  Although the cosmological standard model fits well with measurements~\cite{0067-0049-170-2-377,tegmark2006cosmological}, it raises the pressing question: what causes the accelerated expansion of the universe? To explain this question, researchers have taken two approaches: one is to add dark matter/energy term into the right-hand side of Einstein field equations to provide extra source; the other is modifying the gravitational theory to generate the acceleration term. Although the first approach is a general method, what is the physics of the postulated dark matter and dark energy remains an unsolved mystery. In this paper, we apply the second approach to explain the accelerated expansion of universe. Many methods about modifying gravitational theory have been proposed, such as Lovelock gravity~\cite{lovelock1972four}, Eddington-inspired Born-Infeld gravity~\cite{banados2010eddington}, Horava-Lifshitz gravity~\cite{hovrava2009quantum}, braneworld model~\cite{Maartens2010}, the scalar-tenosr theory~\cite{brans1961mach,fujii2003scalar}, $f(R)$ theory~\cite{DeFelice2010}, $f(R,T)$ theory~\cite{PhysRevD.84.024020}, etc. Thereinto, the scalar-tensor theory was conceived originally by Jordan who started to embed a four-dimensional curved manifold in five-dimensional flat space-time. He pointed out that a constraint in formulating projective geometry can be a four-dimensional scalar field, which enables one to describe a time dependent gravitational ``constant''~\cite{10.2307/97105}. $f(R)$ theory is an extensively studied generalization of General Relativity (GR) that involves modifying the Einstein-Hilbert Lagrangian in the simplest possible way, replacing $R-2\Lambda$ by a more general function $f(R)$~\cite{carroll2004cosmic,dolgov2003can,dick2004newtonian}. After $f(R)$ theory was put forward, researchers found the scalar-tensor theory has a connection with $f(R)$ theory~\cite{fujii2003scalar,PhysRevD.76.063505}. In $2011$, Harko et al.~\cite{PhysRevD.84.024020} generalized $f(R)$ gravity to $f(R,T)$ gravity by introducing an arbitrary function of the Ricci scalar $R$ and the trace of the stress-energy tensor $T$. $f(R,T)$ gravity implies the coupling between matter and geometry that leads to the nongeodesic motion of test particles and an extra acceleration~\cite{jamil2012reconstruction}. Inspired by the relation between the scalar-tensor theory and $f(R)$ theory, we generalize $f(R,T)$ gravity into the two-scalar theory.
\par
  In cosmology, generally, people consider the cosmic content exists in the perfect-fluid form and the cosmological metric is Robertson-Walker metric (RW metric). The general way to discuss the problem of the accelerated expansion is to apply Friedmann equations to obtain the restriction of the cosmic matter's Equation-of-State (EoS) parameter $w$. In this way, one should consider two conditions from Friedmann equations: $\ddot{a}>0$ and $\dot{a}>0$. However, it is complex to derive the restriction of $w$ from $\dot{a}>0$. The other condition, $\ddot{a}>0$, doesn't necessarily mean that universe is accelerated expanding, and it contains two possible results: the accelerated expanding universe and the decelerated contracting universe. In order to obtain the accelerated expanding condition of universe, we need to obtain the expanding condition of universe independent of the restriction from $\ddot{a}>0$ rather than $\dot{a}>0$.
\par
  Raychaudhuri equation can not only estimate whether universe would generate the singularity after infinite time but also reflect the deformation of universe~\cite{poisson2004relativist}. For a congruence of timelike geodesics, Raychaudhuri equation can describe the deformation of the spacelike hypersurface that is orthogonal to this congruence. Because universe locates on a spacelike hypersurface, the deformation of spacelike hypersurface can represent the cosmic expansion or contraction. In GR, Raychaudhuri equation is always negative that means universe is contracting all the time, which generates the singularity after infinite time. However, in modified gravitational theories, Raychaudhuri equation has the possibility to be positive, which means universe is not always contracting and this singularity vanishes. Hence, Raychaudhuri equation can give the expanding condition of universe in modified gravitational theories.
\par
  This paper is organized as follows. In Sec.\uppercase\expandafter{\romannumeral2}, we generalized $f(R,T)$ gravity into the two-scalar theory and derive the field equations of the two-scalar theory. In Sec.\uppercase\expandafter{\romannumeral3}, we reviewed the Raychaudhuri equation in order to obtain the expanding condition of universe in our theory. We discussed the evolution of universe when $f(\chi,\xi)=\chi+\alpha \xi$ in the two-scalar theory in Sec.\uppercase\expandafter{\romannumeral4}. Conclusions and discussion are given in Sec.\uppercase\expandafter{\romannumeral5}.

  \section{THE TWO-SCALAR THEORY FROM $f(R,T)$ THEORY}

  Firstly, we review the connection between $f(R)$ theory and Brans-Dicke (BD) theory~\cite{DeFelice2010}. One considered the following action with a new field $\chi$,
  \begin{equation}
  \begin{aligned}
  \label{eq:The new action of f(R)}
  S=\frac{1}{2\kappa^{2}}\int d^{4}x\sqrt{-g}[f(\chi)+f_{,\chi}(\chi)(R-\chi)]+\int d^{4}x\mathcal{L}_{m}(g_{\mu\nu},\Psi_{M}).
  \end{aligned}
  \end{equation}
  Varying this action with respect to $\chi$, one can obtain
  \begin{equation}
  \begin{aligned}
  \label{eq: vary to X}
  f_{,\chi\chi}(\chi)(R-\chi)=0.
  \end{aligned}
  \end{equation}
  Provided $f_{,\chi\chi}(\chi)\neq0$ it follows that $\chi=R$. Hence the action \eqref{eq:The new action of f(R)} recovers the $f(R)$ action. One defined $\varphi\equiv f_{,\chi}(\chi)$, the action \eqref{eq:The new action of f(R)} can be expressed as
  \begin{equation}
  \begin{aligned}
  \label{eq:The scalar action of f(R)}
  S=\int d^{4}x\sqrt{-g}[\frac{1}{2\kappa^{2}}\varphi R+U(\varphi)]+\int d^{4}x\mathcal{L}_{m}(g_{\mu\nu},\Psi_{M}),
  \end{aligned}
  \end{equation}
  where $U(\varphi)$ is a scalar potential given by $U(\varphi)=\frac{f(\chi(\varphi))-\chi(\varphi)\varphi}{2\kappa^{2}}$. Eq.\eqref{eq:The scalar action of f(R)} is equivalent to Brans-Dicke theory when the BD parameter vanishes~\cite{DeFelice2010}, which shows that $f(R)$ theory can be regarded as a special case of BD theory. Inspired by this method, we intend to generalize $f(R,T)$ theory into a new scalar theory. Let us start from the general 4-dimensional action of $f(R,T)$ theory~\cite{PhysRevD.84.024020}:
  \begin{equation}
  \begin{aligned}
  \label{eq:The f(R,T) action}
  S=\frac{1}{2\kappa^{2}}\int d^{4}x\sqrt{-g}[f(R,T)+2\kappa^{2}\mathcal{L}_{m}(g_{\mu\nu},\Psi_{M})],
  \end{aligned}
  \end{equation}
  where $R$ is the scalar curvature and $T$ is the trace of the stress-energy tensor of matter. $g$ is the determinant of the metric $g_{\mu\nu}$, and $\mathcal{L}_{m}$ is matter Largrangian that depends on $g_{\mu\nu}$ and matter fields $\Psi_{M}$. In this paper we use the natural system of units with $G=c=1$, so that $\kappa^{2}=8\pi$.
  One defined the stress-energy tensor of matter as~\cite{landau1971classical}
  \begin{equation}
  \begin{aligned}
  \label{eq:The stress-energy tensor defined by Lm}
  T_{\mu\nu}=-\frac{2}{\sqrt{-g}}\frac{\delta(\sqrt{-g}\mathcal{L}_{m})}{\delta g^{\mu\nu}}.
  \end{aligned}
  \end{equation}
\par
  Introducing the auxiliary fields $\chi$ and $\xi$, we assumed the action \eqref{eq:The f(R,T) action} as follows:
  \begin{equation}
  \begin{aligned}
  \label{eq:The f(R,T) action in auxiliary field form}
  S=\frac{1}{2\kappa^{2}}\int d^{4}x\sqrt{-g}[f(\chi,\xi)+f_{\chi}(\chi,\xi)(R-\chi)+f_{\xi}(\chi,\xi)(T-\xi)+2\kappa^{2}\mathcal{L}_{m}(g_{\mu\nu},\Psi_{M})],
  \end{aligned}
  \end{equation}
  where  $f_{\chi}(\chi,\xi)=\frac{\partial f(\chi,\xi)}{\partial \chi}$ and $f_{\xi}(\chi,\xi)=\frac{\partial f(\chi,\xi)}{\partial \xi}$.
  Varying this action with respect to $\chi$, one can obtain
  \begin{equation}
  \begin{aligned}
  \label{eq:1111}
  f_{\chi\chi}(\chi,\xi)(R-\chi)+f_{\chi\xi}(\chi,\xi)(T-\xi)=0,
  \end{aligned}
  \end{equation}
  where $f_{\chi\chi}(\chi,\xi)=\frac{\partial^{2} f(\chi,\xi)}{\partial \chi^{2}}$ and $f_{\chi\xi}(\chi,\xi)=\frac{\partial}{\partial \chi}(\frac{\partial f(\chi,\xi)}{\partial\xi})$.
  Making the variation with respect to $\xi$, we obtain
  \begin{equation}
  \begin{aligned}
  \label{eq:2222}
  f_{\xi\chi}(\chi,\xi)(R-\chi)+f_{\xi\xi}(\chi,\xi)(T-\xi)=0.
  \end{aligned}
  \end{equation}
  Eq.\eqref{eq:1111} and Eq.\eqref{eq:2222} show that when
  \begin{equation}
  \begin{aligned}
  \label{eq:0001}
  f_{\chi\chi}(\chi,\xi)f_{\xi\xi}(\chi,\xi)-f_{\chi\xi}^{2}(\chi,\xi)\neq0,
  \end{aligned}
  \end{equation}
  it follows that $R=\chi$ and $T=\xi$, which means the action \eqref{eq:The f(R,T) action in auxiliary field form} is equivalent to the $f(R,T)$ action. However, if the form of $f(\chi,\xi)$ doesn't satisfy the condition \eqref{eq:0001}, this two-scalar action might provide some new physics different than $f(R,T)$ theory.
\par
  On the other hand, the action \eqref{eq:The f(R,T) action in auxiliary field form} can be rewritten into
  \begin{equation}
  \begin{aligned}
  \label{eq:The biscalar tensor model}
  S=&\int d^{4}x\sqrt{-g}\{\frac{1}{2\kappa^{2}}f_{\chi}(\chi,\xi)R+\frac{1}{2\kappa^{2}}f_{\xi}(\chi,\xi)T+U(\chi,\xi)\}\\
  &+\int d^{4}x\sqrt{-g}\mathcal{L}_{m}(g_{\mu\nu},\Psi_{M}),
  \end{aligned}
  \end{equation}
  where $U(\chi,\xi)=\frac{1}{2\kappa^{2}}[f(\chi,\xi)-f_{\chi}(\chi,\xi)\chi-f_{\xi}(\chi,\xi)\xi]$ is a scalar potential. Furthermore, $f_{\chi}(\chi,\xi)$ and $f_{\xi}(\chi,\xi)$ can be regarded as two independent scalar fields. Hence, Eq.\eqref{eq:The biscalar tensor model} is a new type of the scalar-tensor theory that includes two scalar fields, called two-scalar theory. When $f(\chi,\xi)\equiv f(\chi)$, Eq.\eqref{eq:The biscalar tensor model} can degenerate to the Brans-Dicke theory in the situation of the BD parameter $\omega_{BD}=0$\cite{brans1961mach} which is equivalent to $f(R)$ theory~\cite{DeFelice2010}.
\par
  In this paper, we consider $f_{\chi}(\chi,\xi)>0$ in order to rescale the metric by conformal transformation,
  \begin{equation}
  \begin{aligned}
  \label{eq:TO rescale the metric}
  g_{\mu\nu}=\frac{1}{f_{\chi}(\chi,\xi)}\tilde{g}_{\mu\nu}.
  \end{aligned}
  \end{equation}
  It is noteworthy that we define $\tilde{T}_{\mu\nu}\equiv-\frac{2}{\sqrt{-\tilde{g}}}\frac{\delta(\sqrt{-\tilde{g}}\mathcal{L}_{m})}{\delta \tilde{g}^{\mu\nu}}$ as same as the definition of $T_{\mu\nu}$. Based on the definition of the stress-energy tensor, one can prove that its trace is invariant under conformal transformation, $\tilde{T}=T$.
  Via the conformal transformation \eqref{eq:TO rescale the metric}, we can transform the action in Jordan frame \eqref{eq:The biscalar tensor model} into Einstein frame's, given by
  \begin{equation}
  \begin{aligned}
  \label{eq:The f(R,T) action in Einstein frame}
  S=\frac{1}{2\kappa^{2}}&\int d^{4}x\sqrt{-\tilde{g}}[\tilde{R}-\frac{3}{2}\frac{1}{f_{\chi}^{2}(\chi,\xi)}\tilde{g}^{\alpha\beta}\partial_{\alpha}f_{\chi}(\chi,\xi)\partial_{\beta}f_{\chi}(\chi,\xi)\\
  &-V(\chi,\xi)+\frac{f_{\xi}(\chi,\xi)}{f^{2}_{\chi}(\chi,\xi)}\tilde{T}+2\kappa^{2}\frac{1}{f_{\chi}(\chi,\xi)}\mathcal{L}_{m}],
  \end{aligned}
  \end{equation}
  where $V(\chi,\xi)=\frac{\chi}{f_{\chi}(\chi,\xi)}+\frac{f_{\xi}(\chi,\xi)}{f^{2}_{\chi}(\chi,\xi)}\xi-\frac{f(\chi,\xi)}{f^{2}_{\chi}(\chi,\xi)}$ that is a scalar potential.
  Here $\tilde{R}$ is the scalar curvature given by $\tilde{g}_{\mu\nu}$. There are two coupling terms between the scalar fields and matter of universe, `` $\frac{f_{\xi}(\chi,\xi)}{f^{2}_{\chi}(\chi,\xi)}\tilde{T}$ '' and `` $2\kappa^{2}\frac{1}{f_{\chi}(\chi,\xi)}\mathcal{L}_{m}$ ''.
\par
  By the variation of the action \eqref{eq:The f(R,T) action in Einstein frame} with respect to the metric $\tilde{g}^{\mu\nu}$, we obtain the field equations,
  \begin{equation}
  \begin{aligned}
  \label{eq:The Einstein equation in f(R,T) theory}
  \tilde{R}_{\mu\nu}-\frac{1}{2}\tilde{g}_{\mu\nu}\tilde{R}=&\;\frac{f_{\xi}(\chi,\xi)+\kappa^{2}}{f^{2}_{\chi}(\chi,\xi)}\cdot\tilde{T}_{\mu\nu}
  +\frac{f_{\xi}(\chi,\xi)}{f^{2}_{\chi}(\chi,\xi)}\cdot\frac{1}{2}\tilde{T}\tilde{g}_{\mu\nu}\\
  &-\frac{1}{2}V(\chi,\xi)\tilde{g}_{\mu\nu}-\frac{f_{\xi}(\chi,\xi)}{f^{2}_{\chi}(\chi,\xi)}\mathcal{L}_{m}\tilde{g}_{\mu\nu}\\
  &-\frac{3}{4}\frac{1}{f^{2}_{\chi}(\chi,\xi)}\cdot\tilde{g}^{\alpha\beta}\partial_{\alpha}f_{\chi}(\chi,\xi)\partial_{\beta}f_{\chi}(\chi,\xi)\cdot\tilde{g}_{\mu\nu}\\
  &+\frac{3}{2}\frac{1}{f^{2}_{\chi}(\chi,\xi)}\partial_{\mu}f_{\chi}(\chi,\xi)\partial_{\nu}f_{\chi}(\chi,\xi),
  \end{aligned}
  \end{equation}
  where we used the hypothesis of $\frac{\delta(\frac{\delta\mathcal{L}_{m}}{\delta\tilde{g}^{\alpha\beta}})}{\delta\tilde{g}^{\mu\nu}}=0$.
  When $f(\chi,\xi)\equiv f(\chi)$, Eq.\eqref{eq:The Einstein equation in f(R,T) theory} can degenerate to field equations in Einstein frame of $f(R)$ gravity.

  \section{RAYCHAUDHURI EQUATION}

   Raychaudhuri equation can describe the deformation of hypersurface that is orthogonal to a congruence of geodesics. Because universe locates on a spacelike hypersurface, the deformation of hypersurface can be regarded as the evolution of universe. Depending on the Raychaudhuri equation and strong energy condition (SEC), we can obtain the expanding condition of universe. The Raychaudhuri equation for a congruence of timelike geodesics with its tangent vector $u^{\mu}\equiv\frac{dx^{\mu}}{d\tau}$ is written as
  \begin{equation}
  \begin{aligned}
  \label{eq:The Raychaudhuri equation}
  \frac{d\theta}{d\tau}=-\frac{\theta^{2}}{3}-\sigma_{\mu\nu}\sigma^{\mu\nu}+\omega_{\mu\nu}\omega^{\mu\nu}-R_{\mu\nu}u^{\mu}u^{\nu},
  \end{aligned}
  \end{equation}
  where $\theta$, $\sigma_{\mu\nu}$, $\omega_{\mu\nu}$ are respectively the expansion, shear and twist of the congruence of geodesics, and $\tau$ is the proper time of an observer moving along a geodesic. What's more, the Frobenius's theorem~\cite{poisson2004relativist} shows that $\omega_{\mu\nu}=0$ when the congruence of curves (timelike, spacelike, or null) is hypersurface orthogonal. So for this kind of congruence of curves, Eq.\eqref{eq:The Raychaudhuri equation} can be simplified into
  \begin{equation}
  \begin{aligned}
  \label{eq:The simplified Raychaudhuri equation}
  \frac{d\theta}{d\tau}=-\frac{\theta^{2}}{3}-\sigma_{\mu\nu}\sigma^{\mu\nu}-R_{\mu\nu}u^{\mu}u^{\nu}.
  \end{aligned}
  \end{equation}
  In GR, SEC ($(T_{\mu\nu}-\frac{1}{2}Tg_{\mu\nu})u^{\mu}u^{\nu}>0$) can lead to $R_{\mu\nu}u^{\mu}u^{\nu}>0$ that gives rise to $\frac{d\theta}{d\tau}<0$, which means universe is always contracting and the singularity will appear after infinite time, called the focusing theorem~\cite{poisson2004relativist}. Positive contributions from the spacetime geometry to the Raychaudhuri equation are usually interpreted as violation of SEC or the null energy conditions requirements~\cite{poisson2004relativist}. However, in $f(R)$ gravity, one has proved that there may be a positive contribution to Raychaudhuri equation even though SEC holds, which gives the possibility of the repulsive character to expand universe~\cite{santos2017strong}. Inspired by this, we discuss the Raychaudhuri equation in the two-scalar theory, and prove that the positive contribution to Raychaudhuri equation is also existent in this theory when SEC holds, which may make universe expanding.
\par
  As is well-known, in Raychaudhuri equation, ``$-R_{\mu\nu}u^{\mu}u^{\nu}$'' has the possibility to be positive, which gives the positive contribution for Raychaudhuri equation. Based on Eq.\eqref{eq:The Einstein equation in f(R,T) theory}, this term can be rewritten into
  \begin{equation}
  \begin{aligned}
  \label{eq:The expression of Mu}
  M_{\tilde{u}^{\mu}}=&-\tilde{R}_{\mu\nu}\tilde{u}^{\mu}\tilde{u}^{\nu}\\
  =&-\frac{f_{\xi}(\chi,\xi)+\kappa^{2}}{f^{2}_{\chi}(\chi,\xi)}\cdot(\tilde{T}_{\mu\nu}-\frac{1}{2}\tilde{T}\tilde{g}_{\mu\nu})\cdot\tilde{u}^{\mu}\tilde{u}^{\nu}
  -\frac{f_{\xi}(\chi,\xi)}{f^{2}_{\chi}(\chi,\xi)}\cdot\frac{1}{2}\tilde{T}\\
  &-\frac{3}{2}\frac{1}{f^{2}_{\chi}(\chi,\xi)}\partial_{\mu}f_{\chi}(\chi,\xi)\partial_{\nu}f_{\chi}(\chi,\xi)\tilde{u}^{\mu}\tilde{u}^{\nu}\\
  &+\frac{1}{2}V(\chi,\xi)+\frac{f_{\xi}(\chi,\xi)}{f^{2}_{\chi}(\chi,\xi)}\cdot\mathcal{L}_{m},
  \end{aligned}
  \end{equation}
  where $\tilde{u}^{\mu}$ is a timelike vector.
\par
  In the next section, we will discuss $M_{\tilde{u}^{\mu}}$ in the detailed case in order to find the expanding condition of universe.

  \section{THE ACCELERATED EXPANDING CONDITION OF UNIVERSE}

  \subsection{The accelerated condition of scale factor from Friedmann equations}

 Generally, the general way to study the accelerated expansion of universe is to discuss the behavior of the scale factor $\tilde{a}(\tilde{t})$ based on Friedmann equations. In this paper, we derive the Friedmann equations under Einstein frame in two-scalar theory.
\par
 By the scale transformation $g_{\mu\nu}=\frac{1}{f_{\chi}(\chi,\xi)}\tilde{g}_{\mu\nu}$, the RW metric is transformed into
 \begin{equation}
 \begin{aligned}
 \label{eq:The new RW metric}
 d\tilde{s}^{2}=-d\tilde{t}^{2}+\frac{\tilde{a}(\tilde{t})^{2}}{1-kr^{2}}dr^{2}+\tilde{a}(\tilde{t})^{2}r^{2}d\theta^{2}+\tilde{a}(\tilde{t})^{2}r^{2}\sin^{2} \theta d\varphi^{2},
 \end{aligned}
 \end{equation}
 where $d\tilde{t}=\sqrt{f_{\chi}(\chi,\xi)}\cdot dt$ and $\tilde{a}(\tilde{t})=\sqrt{f_{\chi}(\chi,\xi)}\cdot a(t(\tilde{t}))$. Here $f(\chi,\xi)$ is just a time-dependent function. Furthermore, we consider the cosmic content is perfect fluid, which means~\cite{PhysRevD.84.024020}
  \begin{equation}
  \begin{aligned}
  \label{eq:The perfect fluid's Lagrangian density and stess-energy tensor}
  \mathcal{L}_{m}&=p,\\
  \tilde{T}_{\mu\nu}&=(\rho+p)\tilde{u}_{\mu}\tilde{u}_{\nu}+p\tilde{g}_{\mu\nu}.
  \end{aligned}
  \end{equation}
  Based on field equations \eqref{eq:The Einstein equation in f(R,T) theory} and the cosmological metric \eqref{eq:The new RW metric}, we derive Friedmann equations, given by
 \begin{equation}
 \begin{aligned}
 \label{eq:The 00 Friedmann equation}
 3\frac{\dot{\tilde{a}}^{2}}{\tilde{a}^{2}}+3\frac{k}{\tilde{a}^{2}}=\frac{\kappa^{2}}{f_{\chi}^{2}(\chi,\xi)}\rho+\frac{1}{2}\frac{f_{\xi}(\chi,\xi)}{f_{\chi}^{2}(\chi,\xi)}(3\rho-p)+\frac{3}{4}\frac{\dot{f}_{\chi}^{2}(\chi,\xi)}{f_{\chi}^{2}(\chi,\xi)}+\frac{1}{2}V(\chi,\xi),
 \end{aligned}
 \end{equation}
 \begin{equation}
 \begin{aligned}
 \label{eq:The 11 Friedmann equation}
 -2\frac{\ddot{\tilde{a}}}{\tilde{a}}-\frac{\dot{\tilde{a}}^{2}}{\tilde{a}^{2}}-\frac{k}{\tilde{a}^{2}}=\frac{\kappa^{2}}{f_{\chi}^{2}(\chi,\xi)}p+\frac{1}{2}\frac{f_{\xi}(\chi,\xi)}{f_{\chi}^{2}(\chi,\xi)}(-\rho+3p)+\frac{3}{4}\frac{\dot{f}_{\chi}^{2}(\chi,\xi)}{f_{\chi}^{2}(\chi,\xi)}-\frac{1}{2}V(\chi,\xi).
 \end{aligned}
 \end{equation}
 The dot denotes a derivative with respect to time $\tilde{t}$.
 Based on Friedmann equations, the accelerated equation of scale factor $\tilde{a}(\tilde{t})$ is obtained,
 \begin{equation}
 \begin{aligned}
 \label{eq:The accelerated equation of uninverse}
 -2\frac{\ddot{\tilde{a}}}{\tilde{a}}=\frac{1}{3}\frac{\kappa^{2}}{f_{\chi}^{2}(\chi,\xi)}(\rho+3p)+\frac{4}{3}\frac{f_{\xi}(\chi,\xi)}{f_{\chi}^{2}(\chi,\xi)}p+\frac{\dot{f}_{\chi}^{2}(\chi,\xi)}{f_{\chi}^{2}(\chi,\xi)}-\frac{1}{3}V(\chi,\xi).
 \end{aligned}
 \end{equation}
\par
 Generally, different forms of $f(\chi,\xi)$ may lead to different results, in this paper, we would like to show a brief discussion by simply assuming $f(\chi,\xi)=\chi+\alpha \xi$ where $\alpha$ is the coupling strength parameter between the matter fields and spacetime geometry. Obviously, this simple form of $f(\chi,\xi)$ doesn't satisfy the condition \eqref{eq:0001}, which means the theory equipped with this form of $f(\chi,\xi)$ might have some new physics different with $f(R,T)$ theory. And Eq.\eqref{eq:The accelerated equation of uninverse} would be simplified as
 \begin{equation}
 \begin{aligned}
 \label{eq:The simplified accelerated equation of uninverse f(A,B)=A+aB}
 \frac{\ddot{\tilde{a}}}{\tilde{a}}=-\frac{4\pi G}{3}\rho\cdot[1+(3+\frac{\alpha}{2\pi})w].
 \end{aligned}
 \end{equation}
 As is well-known, the current cosmological observations show our universe is accelerated expanding. So, we focus on the case of $\frac{\ddot{\tilde{a}}}{\tilde{a}}>0$. However, the condition of $\frac{\ddot{\tilde{a}}}{\tilde{a}}>0$ doesn't necessarily mean universe is accelerated expanding, universe can be decelerated contracting as well. Hence, the condition of $\frac{\ddot{\tilde{a}}}{\tilde{a}}>0$ is just a necessary condition of the accelerated expanding universe, which can give a restriction of the EoS parameter $w$ for the perfect-fluid content. Eq.\eqref{eq:The simplified accelerated equation of uninverse f(A,B)=A+aB} implies this restriction, given by
 \begin{equation}
 \begin{aligned}
 \label{eq:The restricted equation from universe's accelerated expanding}
 w<-\frac{1}{3+\frac{\alpha}{2\pi}}.
 \end{aligned}
 \end{equation}
 In GR, the necessary condition of the accelerated expanding universe is that the EoS parameter $w$ satisfies the condition of $w<-\frac{1}{3}$, and $w$ is fixed all the time. In our theory, the necessary condition to make universe accelerated expanding depends on a coupling strength parameter $\alpha$ between the matter fields and spacetime, and this parameter can change with time. This time-dependent coupling means the variation of the cosmic content's state with time. For convenience, we define the function $y(\alpha)=-\frac{1}{3+\frac{\alpha}{2\pi}}$ that is the upper brim of the $w$ range, and plot the functional image depicted in Figure 1.

  \subsection{The expanding condition of universe from Raychaudhuri equation}

  Eq.\eqref{eq:The expression of Mu} shows there are two situations of $M_{\tilde{u}^{\mu}}$: $f_{\xi}(\chi,\xi)+\kappa^{2}>0$ and $f_{\xi}(\chi,\xi)+\kappa^{2}<0$. The sign of $f_{\xi}(\chi,\xi)+\kappa^{2}$ determines the direction of the inequality about $M_{\tilde{u}^{\mu}}$ that will influence the next discussion. In this paper, we only discuss the first situation.
\par
  When $f_{\xi}(\chi,\xi)+\kappa^{2}>0$ and SEC holds, we obtain
  \begin{equation}
  \begin{aligned}
  \label{eq:The inequation of Mu lesser}
  M_{\tilde{u}^{\mu}}\leq&-\frac{f_{\xi}(\chi,\xi)}{f^{2}_{\chi}(\chi,\xi)}\cdot\frac{1}{2}\tilde{T}+\frac{1}{2}V(\chi,\xi)+\frac{f_{\xi}(\chi,\xi)}{f^{2}_{\chi}(\chi,\xi)}\cdot\mathcal{L}_{m}\\
  &-\frac{3}{2}\frac{1}{f^{2}_{\chi}(\chi,\xi)}\partial_{\mu}f_{\chi}(\chi,\xi)\partial_{\nu}f_{\chi}(\chi,\xi)\tilde{u}^{\mu}\tilde{u}^{\nu}.
  \end{aligned}
  \end{equation}
\par
  If we take the simple form $f(\chi,\xi)=\chi+\alpha \xi$ as shown above, from the precondition $f_{\xi}(\chi,\xi)+\kappa^{2}>0$, we get the constraint $\alpha>-8\pi$. In this simple form, Eq.\eqref{eq:The inequation of Mu lesser} can be degenerated into
  \begin{equation}
  \begin{aligned}
  \label{eq:The second simplified inequation of Mu lesser}
  M_{\tilde{u}^{\mu}}\leq-\frac{1}{2}\alpha\tilde{T}+\alpha\mathcal{L}_{m}.
  \end{aligned}
  \end{equation}
\par
  Considering the cosmic content as perfect-fluid form, We can obtain
  \begin{equation}
  \begin{aligned}
  \label{eq:The simplified inequation of Mu lesser   perfect fluid}
  M_{\tilde{u}^{\mu}}\leq\frac{1}{2}\alpha\rho(1-w),
  \end{aligned}
  \end{equation}
  where $w=\frac{p}{\rho}$ that is the EoS parameter of the perfect-fluid content of universe. When  $\frac{1}{2}\alpha\rho(1-w)>0$, $M_{\tilde{u}^{\mu}}$ has the possibility to be positive, which includes two situations. The first situation is $w<1$ and $\alpha>0$ and the second situation is $w>1$ and $\alpha<0$, which both are the necessary conditions to make universe expanding. We will discuss these two situations in the following.
\par
  On the other hand, we have assumed that SEC is satisfied in the beginning, which gives a restricted condition about $\rho$ and $p$~\cite{poisson2004relativist},
  \begin{equation}
  \begin{aligned}
  \label{eq:The restricted condition from SEC }
  \rho+3p\geq0,
  \end{aligned}
  \end{equation}
  and it is equivalent to the restriction of the EoS parameter, $w\geq-\frac{1}{3}$.
\par
  As we know, when $M_{\tilde{u}^{\mu}}>0$, the expansion rate of hypersurface $\frac{d\theta}{d\tau}$ has the possibility to be positive. Combining the restricted conditions from $M_{\tilde{u}^{\mu}}>0$ and SEC, we obtain two necessary conditions to generate a expanding universe when $f(\chi,\xi)=\chi+\alpha \xi$:\\
  \begin{equation}
  \begin{aligned}
  \label{eq:The restricted condition from nonsingularity a bigger 0 }
  1.\quad -\frac{1}{3}<w<1 \quad and \quad\alpha>0;\\
  \end{aligned}
  \end{equation}
  \begin{equation}
  \begin{aligned}
  \label{eq:The restricted condition from from nonsingularity a smaller 0 }
  2.\quad w>1 \quad and \quad -8\pi<\alpha<0.
  \end{aligned}
  \end{equation}

 \subsection{The accelerated expanding condition of universe}

 In above we have assumed that the cosmic content exists in the form of perfect fluid. The energy density of the cosmic content is $\rho$ and its pressure is $p$. It's worth noting that $\rho$ could be decomposed into all possible components, $\rho=\Sigma\rho^{(i)}=\rho_{(baryon)}+\rho_{(neutrino)}+\rho_{(radiation)}+\rho_{(dark\,energy)}+
 \rho_{(dark\,matter)}+...$, and the same is true for $p$. In principle, there should be an EoS parameter $w^{(i)}=\frac{p^{(i)}}{\rho^{(i)}}$ associated with each energy component. However, practically, we can regard $w$, the global EoS parameter of the cosmic content, as the weighted average for all relatively dominating components
 \begin{equation}
 \begin{aligned}
 \label{eq: w equation as the weighted average}
 w=\frac{\Sigma p^{(i)}}{\rho},
 \end{aligned}
 \end{equation}
 and thus $w$ varies over cosmic time scale. The variation of $w$ means that the proportion of each matter component changes, which indicates the transformation of matters appears during the evolution of universe.
\par
 When $f(\chi,\xi)=\chi+\alpha \xi$ and $f_{\xi}(\chi,\xi)+\kappa^{2}>0$, we have discussed two restrictions including the accelerated condition of scale factor and the expanding condition of universe. If the scale factor's acceleration is positive, the value range of $w$ should be that $w<-\frac{1}{3+\frac{\alpha}{2\pi}}$, depending on the value of the coupling strength parameter $\alpha$. In addition, if universe is expanding, the value ranges of the coupling strength parameter $\alpha$ and the EoS parameter $w$ should satisfy one of this two situations: (1) $\alpha>0$ and $-\frac{1}{3}<w<1$; (2) $-8\pi<\alpha<0$ and $w>1$. Based on these two restrictions, we find the accelerated expanding condition of universe.
\par
 \begin{figure}[!hbp]
 \centering 
 \includegraphics[scale=0.56]{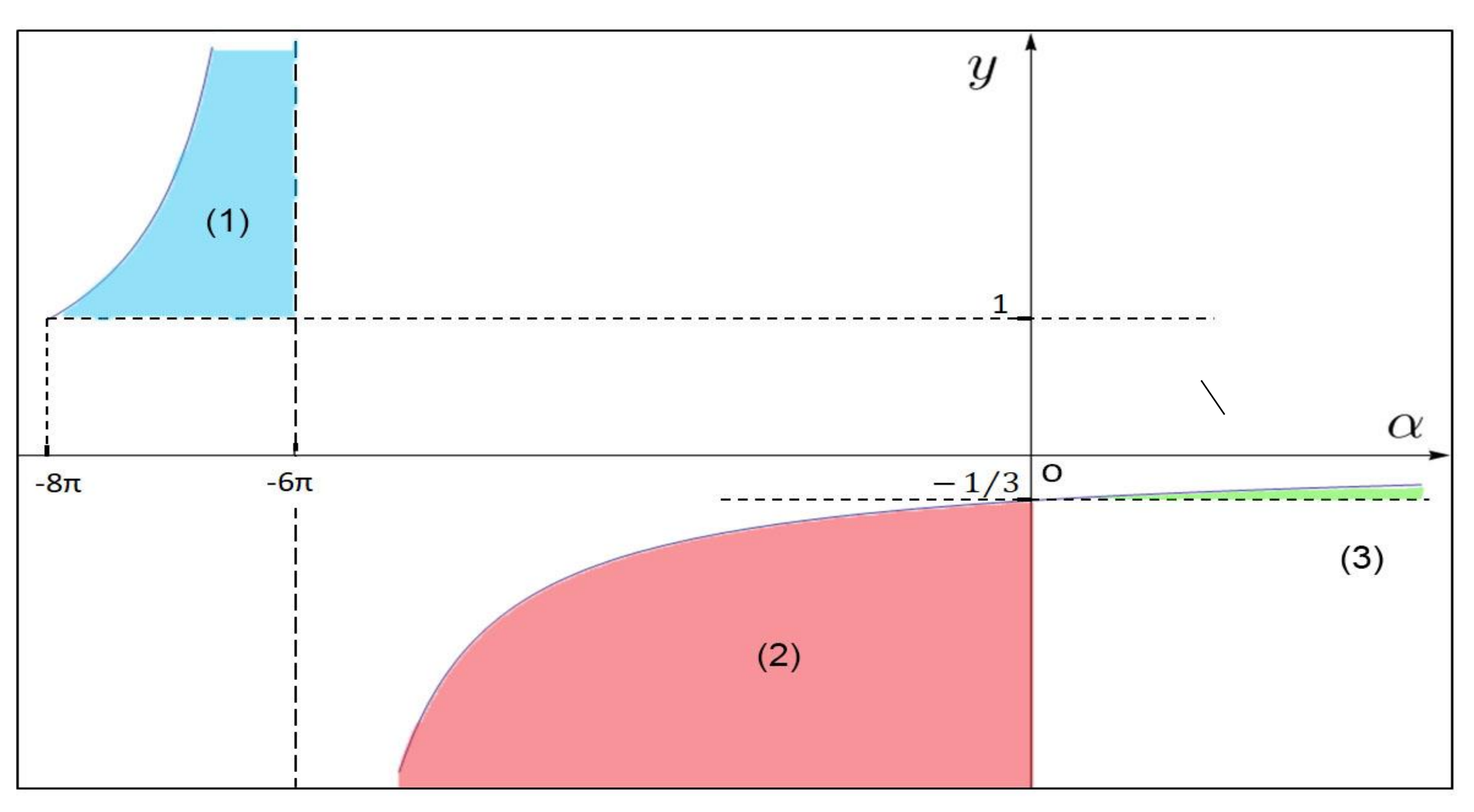}  
 \caption{$y-\alpha$ diagram. $y=-\frac{1}{3+\frac{\alpha}{2\pi}}$ is denoted by the solid dark blue line, which represents the upper brim of the $w$ range. The shaped areas represent the value ranges of $w$ which include the blue one (1), red one (2), and green one (3). The blue area is the first stage ($-8\pi<\alpha<-6\pi$) which represents the early universe; the red area is the second stage ($-6\pi<\alpha<0$) that is an unstable transitional stage of universe; the green area is the third stage ($\alpha>0$) that represents the present universe.}
 \end{figure}
\par
 We obtain two cases in which universe is accelerated expanding. The first case is
 \begin{equation}
 \left\{
 \begin{aligned}
 \label{eq:The w and a range in 1 case}
 &-\frac{1}{3}<w<-\frac{1}{3+\frac{\alpha}{2\pi}},\\
 &\alpha>0,
 \end{aligned}
 \right.
 \end{equation}
 which is corresponding to the green area in Figure 1.
\par
 The second case is
 \begin{equation}
 \left\{
 \begin{aligned}
 \label{eq:}
 &w>1,\\
 &-8\pi<\alpha<-6\pi,
 \end{aligned}
 \right.
 \end{equation}
 which is corresponding to the blue area in Figure 1.
\par
  If we consider the coupling strength parameter $\alpha$ as a monotone increasing function depending on cosmic time, the EoS parameter $w$ will vary with it. And the evolution of universe can be divided into three stages on the basis of the value of $w$. As illustrated in Figure 1, the first stage ($-8\pi<\alpha<-6\pi$), corresponding to the blue area, accords with the first case of the accelerated expanding universe, and it can be regarded as the early universe. The third stage ($\alpha>0$), corresponding to the green area, accords with the second case of the accelerated expanding universe and it can be regarded as the present universe. Between the first and the third stages, the second stage ($-6\pi<\alpha<0$), corresponding to the red area, is decelerated contracting because it satisfies the condition of $\ddot{\tilde{a}}>0$ but violates the expanding condition of universe. Therefore, there is a decelerated contracting state between two accelerated expanding states that corresponds with the early universe and the present universe respectively.
\par
  What's more, we speculate the process of the transformation of cosmic content is as follows. In the very early universe, the content is the original matter whose EoS parameter is equal to $1$. During the first stage, the existence form of content was transformed from the original matter into the existence form of the positive cosmological constant ($\rho=0$, $p\neq0$) by evaporation, which gave rise to the divergency of the upper brim of $w$. At the end of first stage, universe tends to become de Sitter universe. The point of divergence at $\alpha=-6\pi$ between the first and second stage shows the phase transition of the state of universe exists. At the beginning of the second stage, the upper brim of $w$ is much less than $0$, which shows ``exotic field'' energy component~\cite{padmanabhan2003cosmological,peebles2003cosmological} has a huge proportion of the cosmic content. With time elapsing, ``exotic field'' is transformed into other types of matter that have the larger values of the EoS parameter, which leads to the rise of the upper brim of $w$, shown in Figure 1. As time goes on, the value of $\alpha$ goes up and universe evolves naturally from the second stage into the third stage without point of divergence. Between the second stage and the third stage, there is a transition of the state of universe that universe turns into the accelerated expansion from the decelerated contraction. In the third stage, the value range of $w$ tends to $(-\frac{1}{3},0)$ for indefinitely long times, which means the proportion of the dark matter that has the negative value of $w$ would decrease with time. And it also means that, at the end of universe, dust would have a huge proportion of the cosmic content. In other words, the present universe may become a dust one as time goes on.

  \section{Conclusions and discussion}

  In the present paper, we generalize $f(R,T)$ gravity into the two-scalar theory via the method in Ref.\cite{nojiri2003modified}. We also discuss its field equations in Einstein frame by the conformal transformation. Based on the two-scalar theory, when $f(\chi,\xi)=\chi+\alpha\xi$, we discuss the accelerated expanding characteristic of universe and obtain the evolution process of universe.
\par
  In order to obtain the expanding condition of universe in our theory, we discuss the sign of the Raychaudhuri equation. When $f(\chi,\xi)=\chi+\alpha\xi$, we obtain the condition including two cases to make universe expanding: 1.$-\frac{1}{3}<w<1$, $\alpha>0$; 2.$w>1$, $\alpha<0$. On the other hand, Friedmann equations give us another condition: $w<-\frac{1}{3+\frac{\alpha}{2\pi}}$ which is the acceletrated condition of scale factor $\tilde{a}$. What's more, we consider $\alpha$ as the time-dependent coupling strength parameter. The variation of the coupling strength parameter $\alpha$ means the transformation of matter that leads to the change of the EoS parameter $w$.
\par
  Then we combine these two conditions to discuss the accelerated expanding condition of universe. As shown in Figure 1, according to the upper brim of the EoS parameter $w$, the evolution of universe can be divided into three stages: the early universe ($-8\pi<\alpha<-6\pi$, blue area), the transitional period ($-6\pi<\alpha<0$, red area) and the present universe ($\alpha>0$, green area). The transformation process of the cosmic content is as follow.
\par
  In the very early universe, the content is the original matter which has $w=1$. In the first stage, the original matter evaporated into de Sitter vacuum, meanwhile, universe is accelerated expanding. When universe becomes pure de Sitter universe, the phase transition appears, which makes the cosmic state change into the decelerated contracting state in the second stage. In the beginning of the second stage, the upper brim of $w$ is much less than $0$, which means ``exotic field'' energy component has a huge proportion of the universe's content. With time elapsing, the exotic field was transformed into other types of matter that had the larger value of $w$, which leads to the rise of the upper brim of $w$, shown in Figure 1. As time goes on, universe evolves naturally from the decelerated contraction into the accelerated expansion at the critical point $\alpha=0$. In the third stage (the present universe), the upper brim of $w$ tends to $0$ for indefinitely long times, which means the proportion of the dark matter would decrease with time and more dust would be produced out over time. In other words, the present universe would become a dust one with time elapsing.

  \section*{ACKNOWLEDGMENTS}

  We would like to thank the National Natural Science Foundation of China~(Grant No.11571342) for supporting us on this work.

  \bibliographystyle{unsrt}
  \bibliography{reference}

\end{document}